\documentclass[prl,showpacs,amsmath,twocolumn]{revtex4}  
\usepackage{graphicx}

\begin{document}

\title{Exactly Integrable Dynamics of Interface between Ideal Fluid and Light Viscous
Fluid}

\author{Pavel M. Lushnikov$^{1,2}$}

\affiliation{$^1$ Theoretical Division, Los Alamos National
Laboratory,
  MS-B284, Los Alamos, New Mexico, 87545
  \\
  $^2$ Landau Institute for Theoretical Physics, 2 Kosygin Str.,
  Moscow, 119334, Russia
  }
\email{lushnikov@cnls.lanl.gov}


\begin{abstract}
It is shown that dynamics of the interface between ideal fluid and
light viscous fluid is exactly integrable in the approximation of
small surface slopes for two-dimensional flow.  Stokes flow  of
viscous fluid provides a relation between normal velocity and
pressure at interface.  Surface elevation and velocity potential
of ideal fluid are determined from two complex Burgers equations
corresponding to analytical continuation of velocity potential at
the interface into upper and lower complex half planes,
respectively. The interface loses its smoothness if complex
singularities (poles) reach the interface.
\end{abstract}

\pacs{ 47.10.+g, 47.15.Hg, 47.20.Ma, 92.10.Kp}

\maketitle



Dynamics of an interface between two incompressible fluids is an
important fundamental problem which has numerous applications
ranging from interaction between see and atmosphere to flow
through porous media and superfluids. If one neglects gravity and
surface tension, that problem can be effectively solved in some
particular cases in two dimensions with the use of complex
variables. Integrable cases include Stokes flow of viscous fluid
with free surface \cite{Richardson1968}, dynamics of free surface
of ideal fluid with infinite depth
\cite{KuznetsovSpektorZakharov1994}  and finite depth
\cite{DyachenkoZakharovKuznetsov1996}, dynamics of an interface
between two ideal fluids \cite{KuznetsovSpektorZakharov1993},
ideal fluid pushed through viscous fluid in a narrow gap between
two parallel plates (Hele-Shaw flow)
\cite{Richardson1972,Kadanoff1986,Mineev1998}.

Here a new integrable case is found which corresponds to
two-dimensional motion of the interface between heavy ideal fluid
and light viscous fluid in absence of gravity and capillary
forces. The interface position is given by $z=\eta(x,t)$, where
the first, heavier fluid (indicated by index 1) with the density
$\rho_1$ occupies the region $-\infty<z<\eta(x,t)$ and the
second, lighter fluid (index 2) with the density $\rho_2$
occupies the region $\eta(x,t)<z<\infty$.

Suppose that the kinematic viscosity of the fluid 2, $\nu_2$, is
very large so that fluid's 2 flow has small Reynolds numbers and,
neglecting inertial effect in the Navier-Stokes Eq., one arrives
to the Stokes flow Eq. \cite{landaufluid}:
\begin{equation}\label{stokes1}
\nu_{2}\nabla^2{\bf v}_2-\frac{1}{\rho_{2}}\nabla p_{2}=0, \quad
\nabla\cdot {\bf v}_{2}=0,
\end{equation}
where  ${\bf v}_2\equiv
 (v_{2, \, x},v_{2, \, z})$
is the velocity of the fluid 2, $\nabla=(\partial_x,\partial_z)$,
and $p_2$ is the fluid's 2 pressure (similar physical quantities
for the fluid 1 have index 1 below). Additional assumption
necessary for applicability of Eq. $(\ref{stokes1})$ is a small
density ratio,
\begin{equation}\label{rho1rho2}
\rho_2/\rho_1\ll 1, \quad \rho_1\equiv 1,
\end{equation}
which ensure that the fluid 2 responds very fast to perturbations
of the interface as  inertia of the fluid 2 is very small compare
with fluid's 1 inertia while time dependent perturbations of the
fluid 2 decay very fast due to large viscosity $\nu_2$. According
to Eq. $(\ref{stokes1})$, the response of the fluid 2 to motion of
the interface is static. For any given normal velocity of the
interface, $v_{n}$,  Eq. $(\ref{stokes1})$ allows to determine
the pressure $p_2|_{z=\eta}$ at the interface. In other words, the
fluid 2 adiabatically follows the slow motion of the heavy fluid 1
and Reynolds number of the fluid 2 remains small at all time.

The velocity of the  potential motion of ideal the fluid 1, ${\bf
v}_1=\nabla \phi$, can be found from solution of the Laplace Eq.,
$
 \nabla^2\phi=0,$
which is a consequence of the incompressibility condition,
$\nabla\cdot {\bf v}_{1}=0$, for potential flow. Boundary
conditions at infinity are decaying,
$|{\bf v}_1|,\, p_1 \to 0 \quad \mbox{for} \quad z\to -\infty; \nonumber \\
|{\bf v}_2|,\, p_2 \to 0 \quad \mbox{for} \quad z\to +\infty. $

Motion of the interface is determined from the kinematic boundary
condition of continuity of normal component of fluid velocity
across the interface:
\begin{eqnarray}\label{vn1vn2}
v_n\equiv v_{1 \,n}|_{z=\eta}=v_{2 \,n}|_{z=\eta}= \partial_t
\eta\big [ 1+(\partial_x\eta)^2\big ]^{-1/2},
\end{eqnarray}
where $v_{1(2) \,n}={\bf n}\cdot{\bf v}_{1(2)}$ and $ {\bf
n}=(-\partial_x\eta,1)\big [1+(\partial_x\eta)^2\big ]^{-1/2}$
 is the interface normal vector.

A dynamic boundary condition is a continuity of stress tensor,
$\sigma_{1(2), \,jm}=-p_{1(2)}\delta_{jm}+\sigma_{1(2), \,jm}',$
$ \sigma_{1(2), \,jm}'\equiv\rho_{1(2)}\nu_{1(2)}(\frac{\partial
 v_{1(2), \,m}}{\partial x_j}+\frac{\partial v_{1(2), \,j}}{\partial  x_m})$,
$x_1\equiv x, \ x_2\equiv z$, across the interface:
$n_j\sigma_{1, \,jm}|_{z=\eta}=n_j\sigma_{2, \,jm}|_{z=\eta}$
(repetition of indexes $j,m$ means summation from 1 to 2), which
gives two scalar dynamic boundary conditions:
\begin{eqnarray}\label{dyn1}
p_1|_{z=\eta}=p_2|_{z=\eta}+n_mn_j\sigma'_{2, \,jm}|_{z=\eta}, \nonumber \\
 l_mn_j\sigma_{2, \,jm}'|_{z=\eta}=0,
\end{eqnarray}
where the absence of viscous stress in the ideal fluid 1,
$\nu_1=0,$ is used, $n_{m}, l_m$ are components of the interface
normal vector, ${\bf n},$ and the interface tangential vector, $
{\bf l}=(1,\partial_x\eta)\big [1+(\partial_x\eta)^2 \big
]^{-1/2}.$ The pressure $p_1$ of the fluid 1  at the interface can
be determined from a nonstationary Bernoulli Eq., $ \big
[\partial_t \phi+\frac{1}{2}(\nabla \phi)^2+\frac{p_1}{\rho_1}\big
]\Big|_{z=\eta}=0.$

To obtain a closed expression for  interface dynamics in terms of
fluid's 1 variables only, one can first find an expression for
the pressure at the interface through the normal velocity $v_n$ .

It follows from Eq. $(\ref{stokes1})$ that $ \nabla^2 p_2=0$  and
the Fourier transform over $x$ allows to write the solution of the
Laplace Eq. with the decaying boundary condition at $x \to \infty$
as $ p_{2\, k}(z)=p_{2\, k}(0)\exp(-|k|z)\equiv \int dx \,
p_{2}(x,z)\exp(-ikx)$.

To determine ${\bf v}_2|_{z=\eta}$ one can introduce a shift
operator, $\hat L_2$, defined from series expansion:
 ${\bf v}_{2}(x,z)\big|_{z=\eta}\equiv \hat L_2{\bf
v}_2(x,0)=
\Big[(1+\eta\partial_z+\frac{1}{2}\eta^2\partial^2_z+\ldots){\bf
v}_2(x,z)\Big ]\Big|_{z=0} $
 and use Eq. $(\ref{stokes1})$ to find
 $
v_{2, \, x \,k}(z)=\Big [c_k-i k z\frac{ p_{2\,
k}(0)}{2\rho_2\nu_2|k|} \Big ]\exp(-|k|z),$  $ v_{2, \, z
\,k}(z)=\Big [i \, \mbox{sign}(k)c_k +(|k|z+1)\frac{p_{2\,
k}(0)}{2\rho_2\nu_2 |k|}\Big ]\exp(-|k|z),$ where $v_{2, \, x \,
k}(z), \ v_{2, \, z \,k}(z)$ are  the Fourier transform over $x$
of the components of the velocity ${\bf v}_2$ and functions $c_k$,
$p_{2\, k}(0)$ should be determined from the dynamic boundary
conditions $(\ref{dyn1})$.

 Operator $\hat L_2$ can be
expressed, using Eq. $(\ref{stokes1})$, in terms of the  operator
$\hat k:$ $\hat L_2=1-\eta \hat k+\frac{1}{2}\eta^2\hat
k^2+\ldots,$ where the integral operator $\hat k$ is an inverse
Fourier transform of $|k|$ and is given by
\begin{eqnarray}\label{kdef}
\hat k=-\frac{\partial }{\partial x}\hat H.
\end{eqnarray}
Here $\hat Hf(x)=\frac{1}{\pi} P.V.
\int^{+\infty}_{-\infty}\frac{f(x')}{x'-x}dx' $ is the Hilbert
transform and $P.V.$ means Cauchy principal value of integral.
$\hat H$ can be also interpreted as a Fourier transform of $i \,
\mbox{sign}(k)$.

In a similar way one can show that $[\partial_x {\bf
v}_{2}(x,z)]\big |_{z=\eta}=\hat L_2 \partial _x {\bf
v}_{2}(x,0),$ $[\partial_z v_{2, \, x}(x,z)]\big |_{z=\eta}=-\hat
L \hat k v_{2, \, x}(x,0)-\frac{1}{2\nu_2\rho_2}\hat L_2\hat
k^{-1}\partial_x p_2(x,0),$ $p_2(x,\eta)=\hat L_2 p_2(x,0)$ and
using kinematic $(\ref{vn1vn2})$ and dynamic $(\ref{dyn1})$
boundary conditions one can find $p_1(x,\eta)$ as a linear
functional of $v_n$. That linear functional can be expressed in a
form of powers series with respect to small parameter
$|\partial_x\eta|$, which has a meaning of typical slope of the
interface inclination relative to the interface undisturbed
(plane) position.

At leading order approximation over small parameter
$|\partial_x\eta|$ one gets: $p_1(x,\eta)=p_2(x,0),$ $v_n=v_{2,
\,z} (x,\eta)$, and, respectively, response of pressure to normal
velocity is given by
\begin{eqnarray}\label{vnp}
 p_1|_{z=\eta}=2\rho_2\nu_2 \hat k v_n.
\end{eqnarray}
In other words, Eq. $(\ref{vnp})$ determines a static response of
the fluid 2 to the motion of the interface.

Eq. $(\ref{vnp})$ together with the kinematic boundary condition
$(\ref{vn1vn2})$ and the Laplace Eq. for the velocity ponetial
$\phi$ completely defines the potential motion of the fluid 1.

Following Zakharov \cite{Zakharov1968}, one can introduce the
surface variable $ \psi(x)\equiv \phi(x,\eta),$ which is the
value of the velocity potential, $\phi(x,z)$, at the interface.
Kinematic boundary condition $(\ref{vn1vn2})$ can be written at
leading order over small parameter $|\partial_x\eta|$ as
\begin{equation}
\label{etavt} \partial_t \eta=-\hat H v,
\end{equation}
where a new function, $v=\partial_x \psi, $ is introduced which
has a meaning of the tangent velocity of the fluid 1 at the
interface.

Similar to the shift operator $ \hat L_2,$ one can define a shift
operator, $ \hat L_1=1+\eta \hat k+\frac{1}{2}\eta^2\hat
k^2+\ldots,$ which corresponds to the harmonic function $\phi$
with vanishing boundary condition $\phi \to 0$ for $z\to -\infty$.
A Fourier transform of $\phi$, $\phi_k(z)=\phi_k(0)\exp(|k|z)$,
allows to find the components of fluid velocity at the interface:
$(\partial_x \phi)\big|_{z=\eta}=\hat L_1\partial_x\phi(x,0)=\hat
L_1\partial_x\hat L_1^{-1}\psi, \ (\partial_z
\phi)\big|_{z=\eta}=\hat L_1\hat k\hat L_1^{-1}\psi $
through surface variables $\eta, \, \psi$.  Time derivative
$\phi_t$  in the nonstationary Bernoulli  can be found from $
\partial_t \psi=\partial _t\phi|_{z=\eta} +
\partial_t \eta\partial_z \phi|_{z=\eta}$ and one gets at
leading order approximation over $|\partial_x\eta|$:
\begin{eqnarray}\label{vtot}
\frac{\partial v}{\partial t}-\frac{1}{2}\partial_x\Big [(\hat H
v)^2-v^2 \Big] =2\nu_2\rho_2\partial_x^2 v.
\end{eqnarray}
Note that Eq. $(\ref{vtot})$ does not include variable $\eta$
which is a peculiar property of lowest perturbation order over
$|\partial_x\eta|$.

Because the surface tension and gravity is neglected here, the
total energy of two fluid equals to total kinetic energy, $K$.
$K$ decays, $ \frac{d K}{dt} \simeq - \frac{\nu_2\rho_2}{\pi}\int
dk |k|^3|\psi_k|^2 <0,$ due to dissipation in the fluid 2. If the
fluid 2 is absent, which corresponds to $\rho_2 =0$, then $K$ is
conserved, $\frac{d K}{dt}=0$, and the motion of the fluid 1 can
be expressed in the standard Hamiltonian form
\cite{Zakharov1968,KuznetsovSpektorZakharov1994}: $\frac{\partial
\eta}{\partial t}=\frac{\delta K}{\delta \psi},$ $\frac{\partial
\psi}{\partial t}=-\frac{\delta K}{\delta \eta}$.

Equations, similar to $(\ref{etavt}),(\ref{vtot})$  can be
derived for three dimensional motion also with the main difference
that the operator $\hat k$ in three dimensions is not given by
$(\ref{kdef})$ but determined from the Fourier transform  of
$|k|$ over two horizontal coordinates. Subsequent analysis is
however restricted to two dimensional fluid motion only.

The real function $v(x)$ can be uniquely represented as a sum of
two complex functions $v^{(+)}$ and $v^{(-)}$, $ v=
[v^{(+)}+v^{(-)}]/2,$ which can be analytically continued from
real axis $x$ into upper and lower complex half-planes,
respectively. The Hilbert transform acts on these functions as
$\hat H v^{(+)}=iv^{(+)}, \quad \hat H v^{(-)}=-iv^{(-)}$
and Eq. $(\ref{vtot})$ splits into two decoupled complex Burgers
Eqs. for $v^{(+)}$ and $v^{(-)}$:
\begin{eqnarray}\label{vtotpm}
\frac{\partial v^{(\pm)}}{\partial t}+v^{(\pm)}\partial_x
v^{(\pm)} =\tilde \nu\partial_x^2 v^{(\pm)},
\end{eqnarray}
where an effective viscosity, $ \tilde \nu= 2\nu_2\rho_2$ is
introduced to make connection with the standard definition of real
Burgers Eq. \cite{ColeHopf}. Similar reduction of
integro-differential Eq. (like Eq. $(\ref{vtot})$) to complex
Burgers Eq. was done in Ref. \cite{Ablowitz1987}.

If the fluid 2 is absent, $\tilde \nu =0$, complex Burgers Eqs.
$(\ref{vtotpm})$ are reduced to  inviscid Burgers Eqs. (the Hopf
Eqs.) which were derived for ideal fluid with free surface in Ref.
\cite{KuznetsovSpektorZakharov1994} (note that definition of
$v^{(\pm)}$ in this Letter differs from similar definition in
Ref. \cite{KuznetsovSpektorZakharov1994} by a factor $1/2$).
While viscosity $\nu_2$ is large enough to make sure that Reynolds
number in the fluid 2, ${\bf R}_2,$ is small, ${\bf R}_2\sim
v_n/(\nu_2 k)\ll 1$ ($k$ is a typical wave vector of surface
perturbation) but effective viscosity $\tilde \nu$ can be small
provided $\rho_2\ll {\bf R}_2 \ll 1$ so that Reynolds number ,
${\bf R},$ in complex Burgers Eq. $(\ref{vtotpm})$ is large,
${\bf R} \sim {\bf R}_2/\rho_2 \gg 1.$

Complex Burgers Eq. is transformed into the complex heat Eq. $
\frac{\partial u^{(\pm)}}{\partial t}=\tilde \nu\partial_x^2
u^{(\pm)} $ via the Cole-Hopf transform \cite{ColeHopf}:
$v^{(\pm)}=-2\tilde \nu\frac{\partial_x u^{(\pm)}}{u^{(\pm)}}. $
Solution of the heat Eq. with initial data $u^{(\pm)}(x,t)\big
|_{t=0}\equiv u_0^{(\pm)}(x)$,
$u^{(\pm)}(x,t)=(4\pi\tilde \nu t)^{-1/2}
\int^{\infty}_{-\infty}dx'\exp\Big[-\frac{(x-x')^2}{4\tilde \nu
t}\Big ]u_0^{(\pm)}(x'),$
is an analytic function in complex $x$ plane for any $t>0$ because
integral of right hand side (rhs) of this Eq. over any closed
contour in complex $x$ plane is zero (Morera's theorem). Then,
according to the Cole-Hopf transform, solution of the complex
Burgers Eq. can have pole singularities corresponding to zeros of
$u^{(\pm)}(x,t)$. Number of zeros, $n(\gamma)$, of
$u^{(\pm)}(x,t)$ (each zero is calculated according to its order)
inside any simple closed contour $\gamma$  equals to
$\frac{1}{2\pi i}\int_\gamma dx \partial_x
u^{(\pm)}(x,t)/u^{(\pm)}(x,t)$. Integration of Eq.
$(\ref{vtotpm})$ over $\gamma$ allows to conclude that
$n(\gamma)$ is conserved as a function of time provided zeros do
not cross $\gamma$. Thus number of zeros in entire complex plane
can only change in time because zero can be created or annihilated
at complex infinity, $x=\infty$, provided $u^{(\pm)}(x,t)$ has an
essential singularity at complex infinity.

From physical point of view it is important that zeros of
$u^{(\pm)}(x,t)$ can reach real axis $x=Re(x)$ which
distinguishes the complex Burgers Eq. from the real Burgers Eq.
Solution of the real Burgers Eq., which corresponds to Eq.
$(\ref{vtotpm})$ with
$v^{(\pm)}(x,t)\big|_{t=0}=Re\big[v^{(\pm)}(x,t)\big|_{t=0}\big]$,
has global existence (remains smooth for any time), while
solution of the complex Burgers generally exists until some zero
of $u^{(\pm)}(x,t)$ hits real axis $x$ for the first time.

To make connection with inviscid case
\cite{KuznetsovSpektorZakharov1994} one can look at initial
condition for $v^{(+)}(x,0)$ with one simple pole in the lower
half-plane:
\begin{eqnarray}\label{simplepole}
v^{(+)}(x,0)=\frac{2A}{x +ia}, \quad Re(a)>0.
\end{eqnarray}
 Solution of the
inviscid ($\tilde \nu=0)$ Burgers Eq. with initial condition
$(\ref{simplepole})$ gives \cite{KuznetsovSpektorZakharov1994}: $
v^{(+)}_{inviscid}(x,t)=\frac{4A}{x+ia+\sqrt{(x+ia)^2-8At}}, $
which has two moving branch points: $ x_{1,2}=-ia\pm 2\sqrt{2At}.$
One of these branch points reaches real axis in a finite time if
either $A<0$ or $Re(A)\neq 0.$ As the branch point touches the
real axis, the inviscid solution is not unique any more and
 the interface looses its smoothness
\cite{KuznetsovSpektorZakharov1994}.

Consider now solution of the viscous Burgers Eq. $(\ref{vtotpm})$
with nonzero effective viscosity $\tilde \nu$ and with the simple
pole conditions $(\ref{simplepole})$.  Respectively, initial
condition  for the heat Eq., is given by $
u_0^{(+)}=(x+ia)^{-A/\tilde \nu}$ and has branch point at
$x=-ia$. Solution of the heat Eq.  gives
 $
u^{(+)}(x,t)=\exp\big(i\frac{\pi \tilde \mu}{2}\big) H_{\tilde
\mu}\Big(-\frac{i}{2\sqrt{\tilde \nu t}}[x+ia]\Big ), $ where
$\tilde \mu\equiv -A/\tilde \nu$ and $H_\mu(z)$ is the Hermite
function defined as $
H_\mu(z)=\frac{2^{\mu+1}}{\sqrt{\pi}}e^{z^2} \int^\infty_0 dy
e^{-y^2}y^{\mu}\cos\big ( 2zy-\frac{\pi\mu}{2} \big ).$ Zeros of
$u^{(+)}(x,t)$ (and, equivalently, poles of $v^{(+)}(x,t)$)  move
in complex $x$ plane with time as (see Figure 1)
\begin{eqnarray}\label{xj}
x_j(t)=i(2\sqrt{\tilde \nu t}z_j-a),
\end{eqnarray}
where $z_1, z_2, \ldots,$ are complex zeros of the Hermite
function.

\begin{figure}
\begin{center}
\includegraphics[width = 3.5 in]{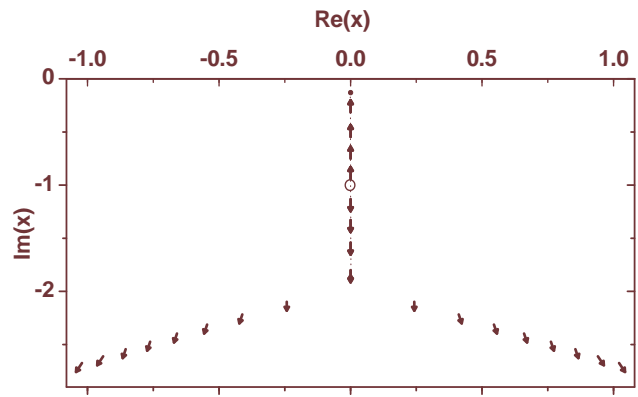}
\caption{Motion of poles of the complex velocity, $v^{(+)}(x,t)$,
in complex $x$ plane for $A=-1/8,\, \tilde \mu=7.6, \, \tilde
\nu\simeq 0.0164, \, a=1, \ t=0.8.$ Arrows point out the position,
direction and magnitude of moving poles.  Uppermost arrow
designates the pole which corresponds to zero of the Hermite
function with the largest real part (that pole first reaches real
axis for $t_{viscous}\simeq 1.91$ producing singularity of the
interface surface). Dotted line connects two branch points
(filled circles) of inviscid solution. Upper branch point reaches
real axis for $t_{inviscid}=1$ which corresponds to
 singularity in the solution of
the inviscid Burgers Eq. Empty circle designates the simple pole
initial condition  $(\ref{simplepole})$. Viscous solution becomes
singular at later time compare to inviscid solution,
$t_{viscous}>t_{inviscid}.$ } \label{fig:fig1}
\end{center}
\end{figure}

Consider a particular case, $ \tilde \mu =n,$  $n$ is a positive
integer number. The Hermite function is reduced to the Hermite
polynomial $H_n(z)$ which has $n$ zeros, $z_1, z_2, \ldots, z_n$
located at real axis $z=Re(z)$, $z_n$ corresponds to the largest
zero. Location of real zeros of the Hermite function with real
$\tilde \mu$ is close to location of zeros of the Hermite
polinomial with the closest integer $n$ to the given $\tilde \mu$
while zeros with nonzero imaginary part (which corresponds to
tails with nonzero real part in Figure 1) disappear for $\tilde
\mu=n.$ Zeros of the Hermite polinomial are moving with time
parallel to imaginary axis $x=Im(x)$ in complex $x$ plane
according to $(\ref{xj})$ and the complex velocity $v^{(+)}$ is
described by set of moving poles:
\begin{eqnarray}\label{vj}
v^{(+)}=-2\tilde \nu\sum\limits_{j=1}^n\frac{1}{x-x_j(t)}.
\end{eqnarray}
$v^{(-)}$ is given by the same expression with $x_j$ replaced by
their conjugated values $\bar x_j$.

Eqs. $(\ref{vtotpm})$ have also another wide class of solutions,
``pole decomposition", corresponding to Eq. $(\ref{vj})$ with
$\frac{d x_j}{dt}=-2\tilde \nu \sum\limits_{l=1, \, j\neq
l}^n\frac{1}{x_j-x_l}$, $n$ is arbitrary positive integer
\cite{Choodnovsky1977}.  Simple pole initial condition
$(\ref{simplepole})$ with $ \tilde \mu =n$ is particular case for
which $x_j|_{t=0}=0$ for any $j$.

As $v(x,t)$ is known from solution of the heat Eq. and the
Cole-Hopf transform one can find $\eta(x,t)$ from Eq.
$(\ref{etavt})$. Interface dynamics is determined from the most
rapid pole of $v^{(\pm)}$ which first reaches real axis,
$x=Re(x)$. E.g., for initial condition $(\ref{simplepole})$, the
pole singularity of $v^{(+)}$ first hits real axis, $x=Re(x)$,
from below at time $t_{viscous}=\frac{Re(a)^2}{4\tilde \nu
Re(z_{max})^2}$, where $z_{max}$ is a complex zero of the Hermite
function with the largest real part for given $\tilde \mu$.
Simultaneously, the pole singularity of $v^{(-)}$ first hits real
axis from above at the same point.
\begin{figure}
\begin{center}
\includegraphics[width = 3.5 in]{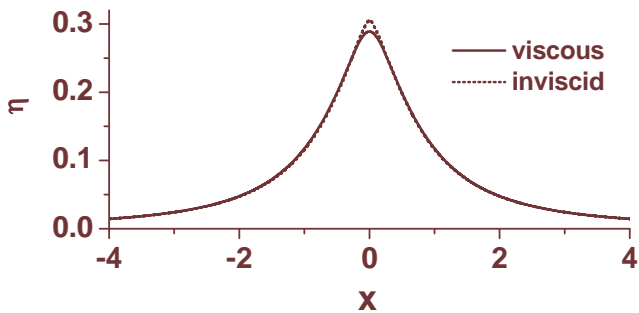}
\caption{ The interface position, $\eta(x,t)$, according to
solution of Eqs. $(\ref{etavt}),(\ref{vtot})$  with finite
viscosity, $\tilde \nu=1/64$ (solid line) and zero viscosity,
$\tilde \nu=0,$ (dotted line) for $A=-1/8,\, \, a=1, \ t=1$.
Viscous solution has 8 moving poles while inviscid solution is
singular at $x=0$ ($\partial^2_{xx}\eta|_{x=0}\to -\infty$ as $t
\to t_{inviscid}=1$). Both solutions are almost indistinguishable
outside a small neighborhood around $x=0$.  As $\tilde \mu$
increases, the viscous solution approaches invisicid. }
 \label{fig:fig2}
\end{center}
\end{figure}
Figure 2 shows $\eta(x)$ at the time, $t=t_{inviscid}$, when
singularity (branch point) of inviscid solution first reaches the
interface breaking analyticity of inviscid solution. It is seen
that viscous solution significantly deviates from inviscid one
only in the narrow domain around $x=0$.

Viscous solution remains analytic for $t>t_{inviscid}$ until
$t<t_{viscous}$ ($t_{viscous}\simeq 1.91$ for parameters in
Fig.1). However, for $t\to t_{viscous}$, surface elevation behaves
as $\eta\simeq (-a/2z_{max}^2)\log{\big [ x^2+(2\sqrt{t\tilde
\nu}z_{max}-a)^2\big ]}$ near $x=0$ (it is set here
$Im(a)=Im(A)=0$) meaning that small slope approximation used for
derivation of Eqs. $(\ref{etavt}),(\ref{vtot})$ is violated for
$t\to t_{viscous}$ and full hydrodynamic Eqs. should be solved
near singularity. One can find a range of applicability of Eqs.
$(\ref{etavt}),(\ref{vtot})$ by looking at correction to these
Eqs. E.g. the analysis  for parameters of Fig. 1 shows that the
correction is important for $t \gtrsim 0.9 t_{viscous}$ (for $t =
0.9 t_{viscous}$ correction to $\eta|_{x=0}$ is about $30\%.)$
Detail consideration of that question is outside the scope of
this Letter. Note that the question whether an actual singularity
of the interface surface occurs in full hydrodynamic Eqs. remains
open.

To make connection with dynamics of ideal fluid with free surface
(corresponds to the inviscid Burgers Eqs.)
\cite{KuznetsovSpektorZakharov1994} one can  consider a limit
$\tilde \nu \to 0$ and, respectively, $\tilde \mu\to \infty$. It
can be shown from the asymptotic analysis of the integral
representation of the Hermite function that the largest zero,
$z_{max},$  is given by $z_{max}= 2^{1/2}\tilde \mu^{1/2}+O(\tilde
\mu^{-1/6})$ The leading order term, $2^{1/2}\tilde \mu^{1/2}$,
exactly corresponds to the position of the upper branch point of
inviscid solution (see Fig. 1) while term $O(\tilde \mu^{-1/6})<0$
is responsible for the difference between $t_{inviscid}$ and
$t_{viscous}$. Even for moderately small $\tilde \nu$ as in Fig.
1 that difference is numerically close to 1 because of small power
$\tilde \mu^{-1/6}$.

It is easy to derive a wide class of initial conditions for which
solution $(\ref{etavt}),(\ref{vtot})$ exists globally and the
interface remains smooth at all times. E.g. one can take
$u^{(+)}=a_0e^{ik_0x-\tilde \nu k_0^2 t}, $ $k_0=Re(k_0)>0$ or any
sum of imaginary exponent which ensure that there is no zeros at
$Im(x)=0$. However, it we suppose that there is a random force
pumping of energy into system (or random initial condition) then
one can expect that some trajectories with nonzero measure would
have poles which reach real axis in a finite time.

In conclusion, one can mention  possible physical applications.
Eqs. $(\ref{etavt}),(\ref{vtot})$ describe a free surface dynamics
of Helium II with both normal ($\nu_2\neq 0$) and superfluid
($\nu_1=0$) components. Derivation of these Eqs. is slightly
different from given in this Letter because both fluids occupy
the same volume but resulting Eqs. are exactly the same as
$(\ref{etavt}),(\ref{vtot})$. For classical fluids viscosity is
nonzero but $\nu_1$ can be neglected and the fluid 1 can be
considered as ideal fluid provided the ratio of dynamic
viscosities of two fluids is large, $\nu_2
\rho_2/(\nu_1\rho_1)\gg 1$. E.g. that ratio is $\sim 900$ for
glycerin and mercury while ratio of their densities is $\sim 0.09$
which makes them good candidates for experimental test of the
analytical result of this Letter.


The author thanks M. Chertkov, I.R. Gabitov, E.A. Kuznetsov, H.A.
Rose, and V.E. Zakharov for helpful discussions.

Support was provided by the Department of Energy, under contract
W-7405-ENG-36.








\end{document}